\documentclass[aps,prb,amsmath,amssymb,titlepage,superscriptaddress,12pt]{revtex4-1}	
\usepackage[english]{babel}
\usepackage{graphicx}
\usepackage{amsmath}
\usepackage{amssymb}
\usepackage{epstopdf}
\usepackage{amsfonts}
\usepackage{bm}
\usepackage{mathdots}
\usepackage{times}
\usepackage{mathtools}
\usepackage{hyperref}

\usepackage{color}

\newcommand{\astcycl}{\mathrlap{\kern0.085em{\circlearrowright}}\ast}
\newcommand{\taucycl}{\mathrlap{\kern0.42em{\bullet}}\circlearrowright}

\begin{document}
\title{Cooling by photo-doping -- \\Light-induced symmetry breaking in the Hubbard model}
\author{Philipp Werner}
\email{corresponding author: philipp.werner@unifr.ch}
\affiliation{Department of Physics, University of Fribourg, 1700 Fribourg, Switzerland}
\author{Martin Eckstein}
\affiliation{Department of Physics, University of Erlangen-N\"urnberg, 91058 Erlangen, Germany}
\author{Markus M\"uller}
\affiliation{Paul Scherrer Institute, Condensed Matter Theory, PSI Villigen, Switzerland}
\author{Gil Refael}
\affiliation{Department of Physics, California Institute of Technology, Pasadena, CA 91125, USA}

\begin{abstract}
\vspace{5mm}
An elusive goal in the field of driven quantum matter is the induction of long-range order. Here, we demonstrate a mechanism based on light-induced evaporative cooling of holes in a correlated electron system. Since the entropy of a filled narrow band grows rapidly with hole doping, the isentropic transfer of holes from a doped Mott insulator to such a band results in a drop of temperature. Strongly correlated Fermi liquids and symmetry-broken states could thus be produced by dipolar excitations. Using nonequilibrium dynamical mean field theory, we show that suitably designed chirped pulses allow to realize this cooling effect. In particular, we demonstrate the emergence of antiferromagnetic order in a system which is initially in a weakly correlated state above the maximum N\'eel temperature. Our work suggests a general strategy for inducing strong correlation phenomena and electronic orders in light-driven materials or periodically modulated atomic gases in optical lattice potentials.
\end{abstract}
\maketitle

Inducing or enhancing electronic orders by some nonequilibrium process is an intriguing prospect, which captured the attention of many scientists \cite{Denny2015,Okamoto2016,Sentef2017,Murakami2017,Babadi2017,Claassen2017,Kennes2017,Mazza2017,Murakami2017exc,Nava2018,Fabrizio2018,Li2018} following the recent observation of an apparent high-temperature superconducting state in light-driven cuprates \cite{Kaiser2014,Hu2014} and fulleride compounds \cite{Mitrano2016}. Some of the theoretical proposals which have been put forward to explain these experiments focus on the cooling of quasi-particles or phase fluctuations in a periodically driven state. In Ref.~\onlinecite{Nava2018} it has been argued that the quasi-particles in phonon-driven K$_3$C$_{60}$ are cooled via the creation of entropy-rich spin-triplet excitons. In an earlier study focusing on bi-layer cuprates \cite{Denny2015}, the authors showed that a time-periodic modulation of the plasma frequency may result in the cooling of the low-energy inter-bi-layer plasmon modes by energy transfer to the high-energy intra-bi-layer plasmon modes. While these proposals are very interesting, they rely so far on relatively simple model calculations which may not fully capture the interaction effects in driven many-body systems. 

Simulations of photo-excited or phonon-driven correlated electron systems \cite{Herrmann2017,Murakami2017} based on nonequilibrium dynamical mean field theory (DMFT) \cite{Aoki2014} have generically produced heating effects and a melting of electronic orders \cite{Werner2012,Golez2016,Li2018}. Moreover, the formation of quasi-particles in photo-doped Mott insulators has been found to be extremely slow \cite{Eckstein2013}, even in the presence of cooling by a phonon bath. While an enhancement of pairing susceptibilities in photo-doped Mott insulators has been reported \cite{Werner2018}, the observed effect was too small to trigger a symmetry-breaking. So far, it thus remained unclear if quasi-particle cooling and electronic ordering transitions can be induced by periodic driving or photo-doping if heating and thermalization effects are accounted for. 

Here we demonstrate effective cooling by optically enabling the evaporation of holes from a system of interest (e.g. a partially filled Hubbard band) into a completely filled, almost flat band. The protocol we envision involves coupling a hole-doped Mott insulator by dipolar excitations to an initially filled core band. Since the entropy increase per hole is large in a full narrow band, a transfer of holes at constant total entropy would result in strong cooling; a feature shared with other isentropic cooling schemes\cite{Bernier2009,Chiu2018,Fabrizio2018}. In the rotating frame, optical driving between the bands induces a tunneling from the Hubbard to the flat band, whose energy is shifted by the driving frequency $\Omega$. The evaporative cooling is most effective if $\Omega$ matches the difference in effective chemical potentials between the two bands. In this case, hot holes are ejected from the Hubbard band in a narrow energy region below the chemical potential, which produces a steeper (and hence colder) distribution. The entropy of the system can be reduced by this mechanism down to a temperature which is essentially limited by the width of the narrow band. Using nonequilibrium DMFT simulations, we show that this idealized scenario can be approximately realized by simple and realistic driving protocols, resulting in optically induced cooling and electronic ordering.

In cold atom experiments the two subsystems could be realized by chains or layers at different potentials. In condensed matter, the full band represents a low-lying ligand band or core states. We will refer to the Hubbard band as the {\it system} and to the full flat band as the {\it core}. Initially, the system is decoupled from the core, both having equal temperature above the maximum N\'eel temperature of the Hubbard model (which is reached at half-filling). A laser pulse then excites holes from the system to the core by dipolar excitations. The interacting system quickly relaxes to a thermal state with an effective temperature $T_\text{eff}$. (While in the rotating frame system and core may look almost equilibrated, the full system in the original frame is in a highly nonthermal state.) To optimize cooling the driving frequency should be instantaneously adjusted to match the evolving chemical potential difference between the bands. As we will show, a linearly chirped pulse suffices to realize a substantial charge transfer and to cool the system far below the N\'eel temperature. 

\begin{figure}[t]
\begin{center}
\includegraphics[angle=-90, width=0.39\columnwidth]{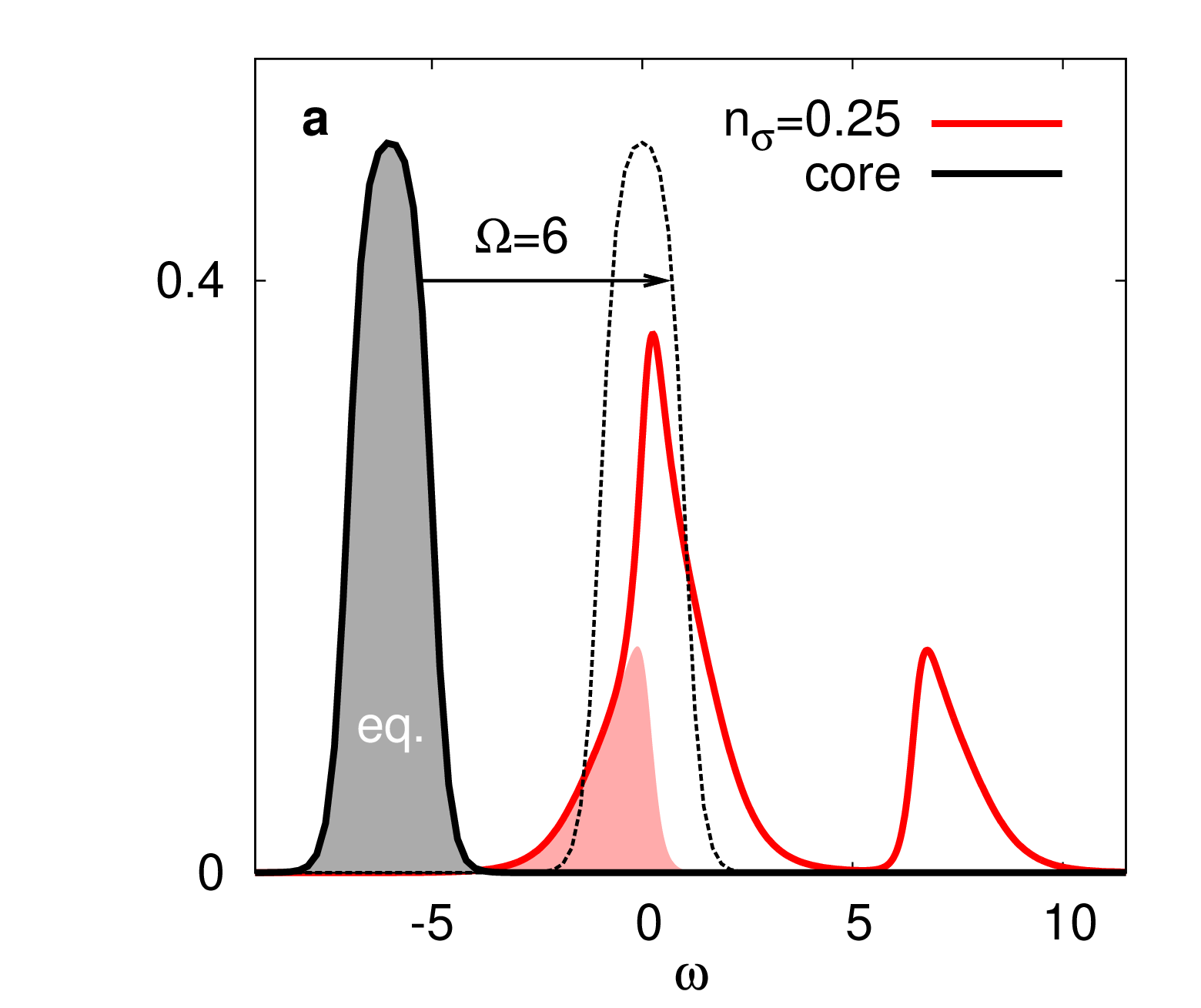}
\hspace{5mm}
\includegraphics[angle=-90, width=0.39\columnwidth]{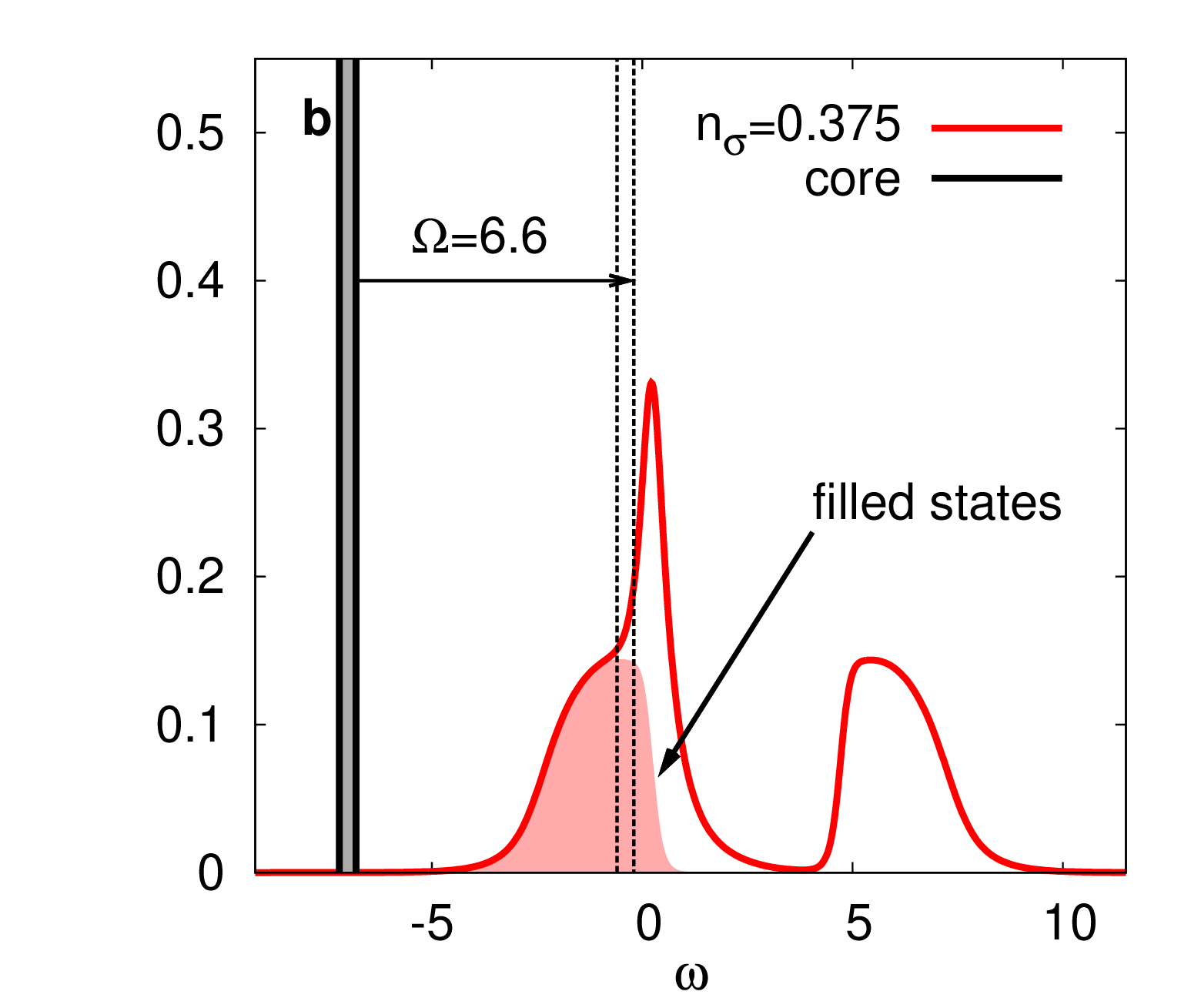}
\includegraphics[angle=-90, width=0.39\columnwidth]{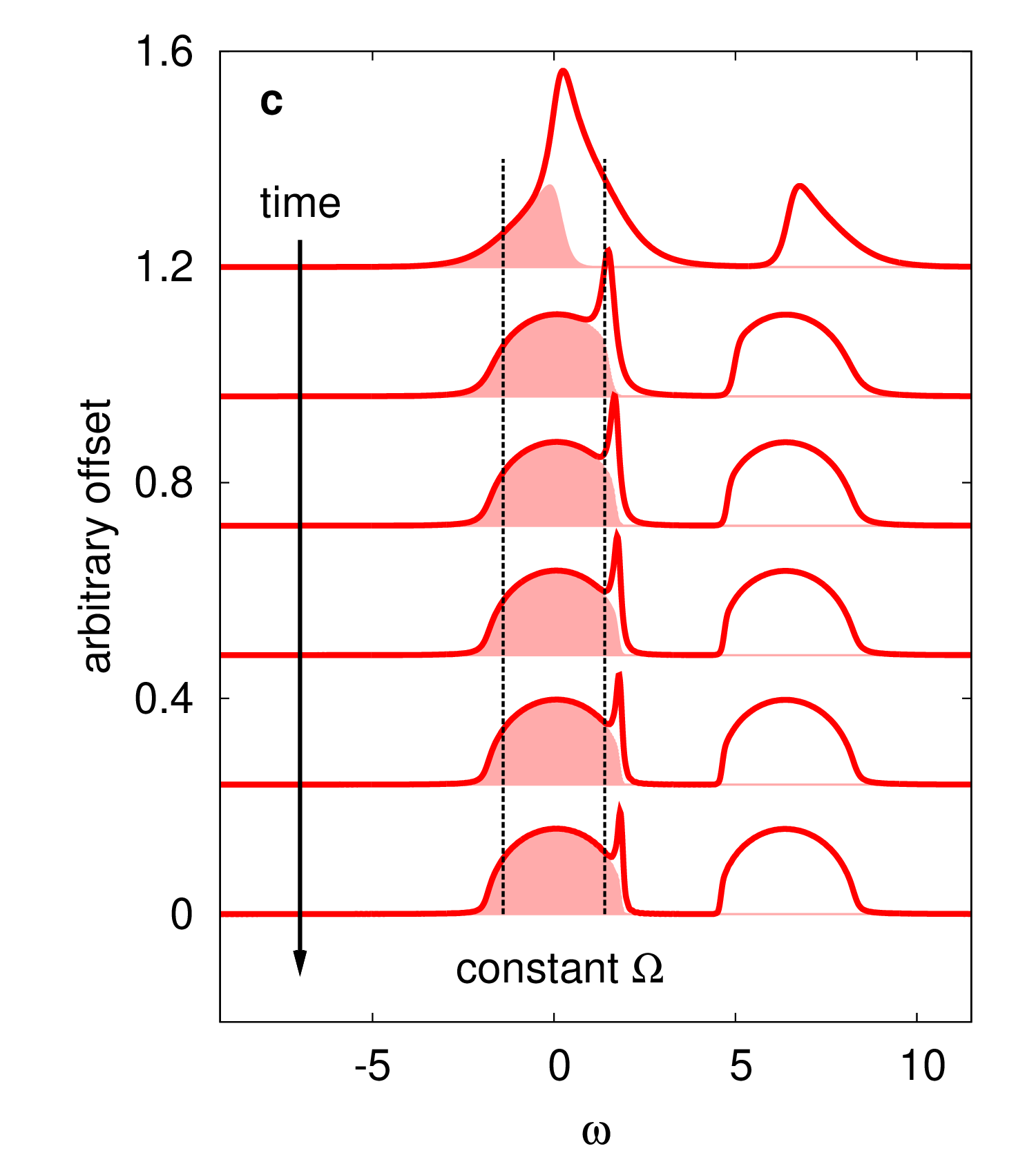}
\hspace{5mm}
\includegraphics[angle=-90, width=0.39\columnwidth]{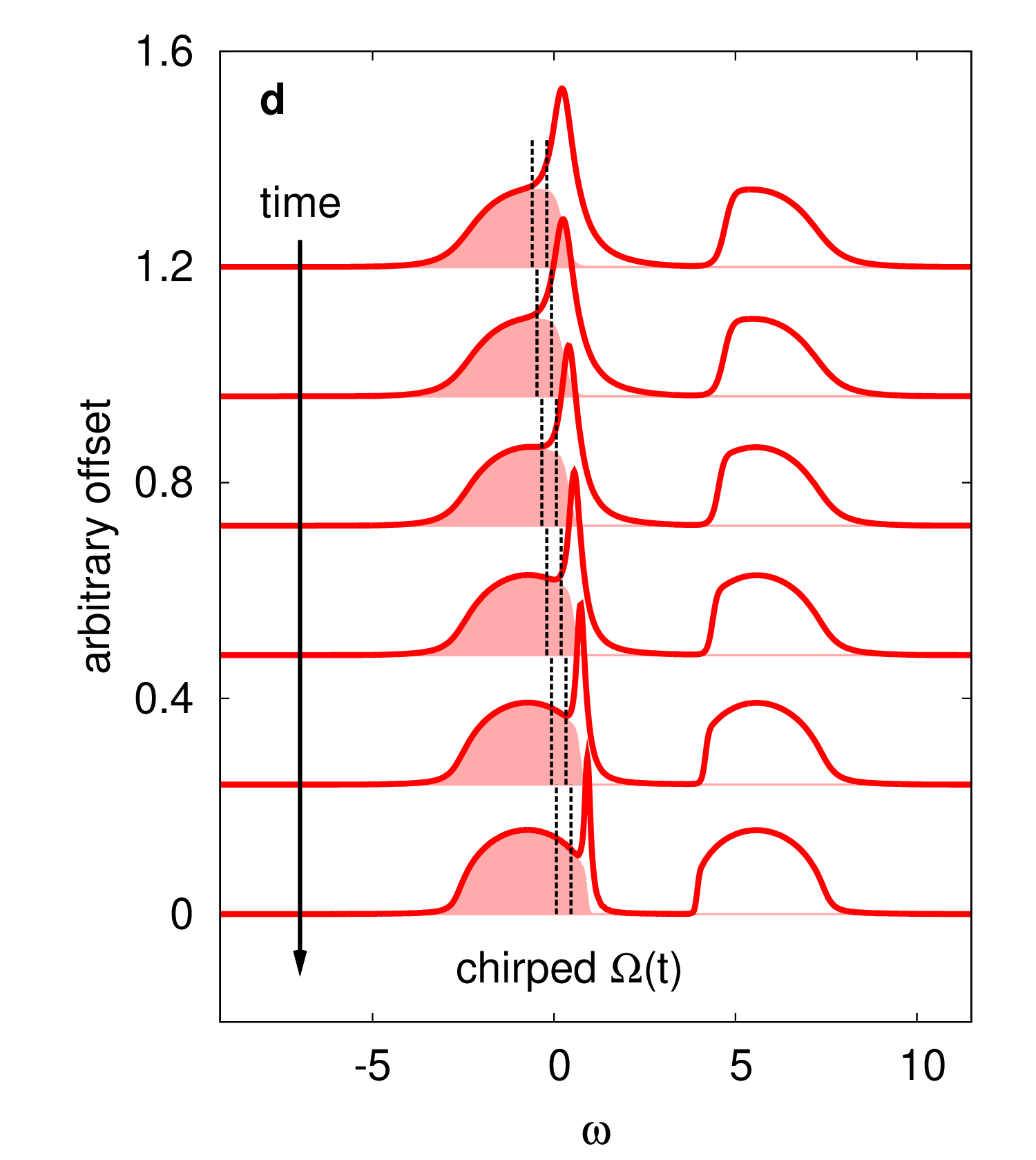}
\caption{
{\bf Spectral functions before and after photo-excitation.} Panel {\bf a} illustrates the initial state in the first set-up: a quarter-filled Hubbard model (red) and core states of bandwidth 2 (black), which are treated as a noninteracting electron bath. 
Panel {\bf b} shows the second set-up: a 3/8 filled Hubbard model (red) and a narrow core band of width 0.4 (black).  In both cases, the interaction parameters are $U_\text{system}=6$, $U_\text{core}=0$, and the initial temperature is $T=0.2$.
Panel {\bf c} illustrates the evolution of the spectral function (lines) and occupation (shaded region) with pulse duration in the first set-up, for constant driving frequency $\Omega=6$ and amplitude $a_\text{max}=0.8$. 
Panel {\bf d} shows analogous results for the second set-up and a chirped pulse with $\Omega_\text{in}=6.6$ and amplitude $a_\text{max}=0.1625$. 
}
\label{fig_illustration}
\end{center}
\end{figure}

We will consider two set-ups, representing the opposite limits of a large number of core bands, and a single core band, respectively. The first set-up (Fig.~\ref{fig_illustration}a,c) consists of a Hubbard model on an infinite-dimensional Bethe lattice, which can be solved exactly using DMFT \cite{Georges1996}, and core levels represented by a noninteracting electron bath, which remains in equilibrium at constant temperature and filling (see Methods). The bandwidth of the noninteracting Hubbard model is $4v$ (corresponding to a hopping $v_\text{system}=v$), that of the core band is $2v$, and we use $v$ ($\hbar/v$) as the unit of energy (time). (For a bare bandwidth of 4 eV, the unit of time is 0.66 fs.) The electrons are transferred from the filled noninteracting core levels (electron bath)  into the lower Hubbard band via a dipolar excitation with frequency $\Omega$, maximum amplitude $a_\text{max}$ and pulse length $\delta$ (see Methods).  

In Fig.~\ref{fig_illustration}a we show the density of states (DOS) of the system (red) and core levels (black) in the initial state, together with their occupations (shaded regions).  The Hubbard interaction $U=6$ is larger than the bandwidth, so that the DOS of the system is split into upper and lower Hubbard bands. We start in a quarter-filled state, with chemical potential in the lower Hubbard band. The dashed black line represents the core level DOS shifted by $\Omega=6$. The initial temperature of the system and core bands is $T=0.2$ (inverse temperature $\beta=5$), which is above the maximum $T_\text{N\'eel}\approx 0.15$ reached for $U=6$ at half-filling\cite{Werner2012}. 

\begin{figure}[t]
\begin{center}
\includegraphics[angle=-90, width=0.39\columnwidth]{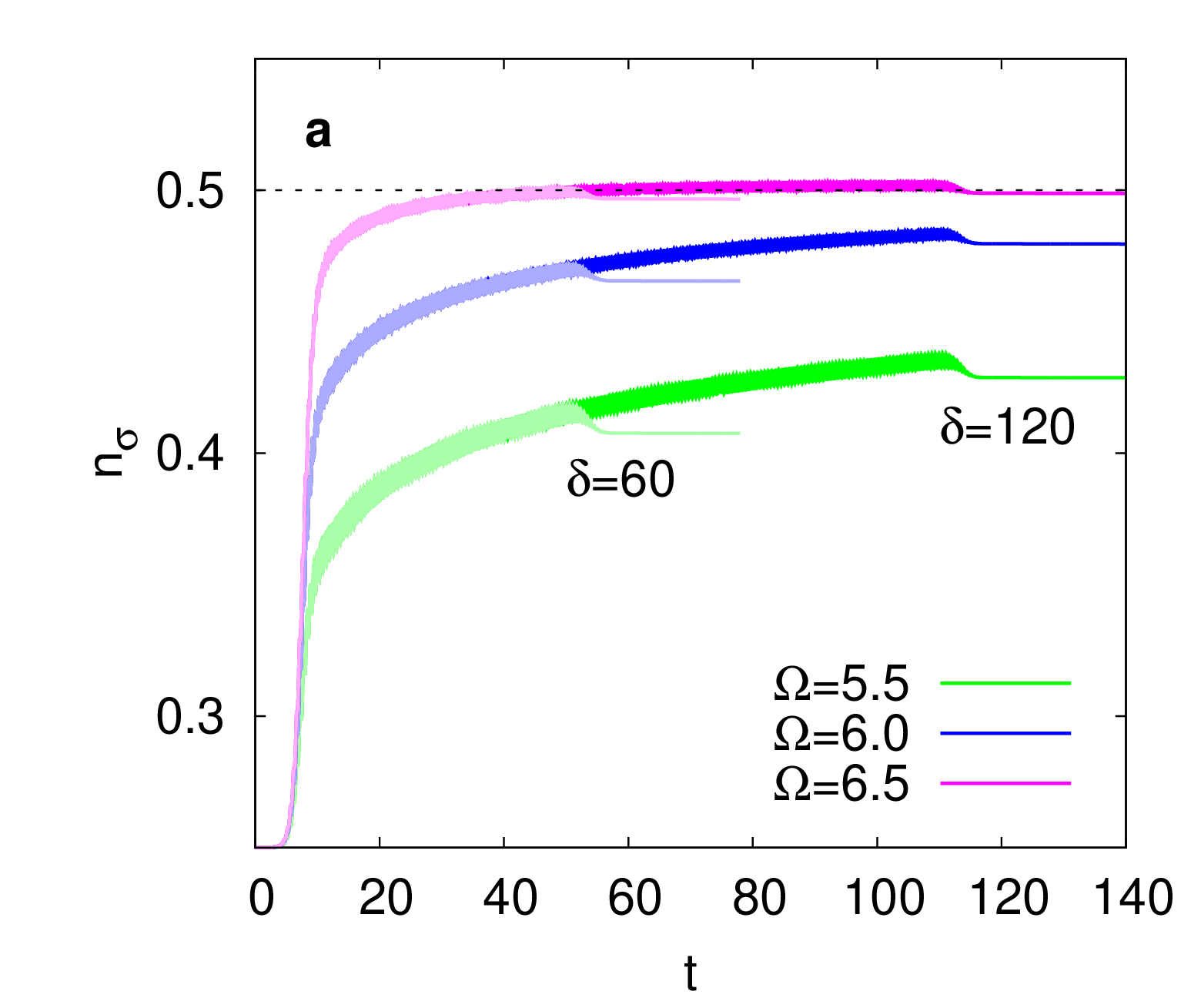}
\hspace{5mm}
\includegraphics[angle=-90, width=0.39\columnwidth]{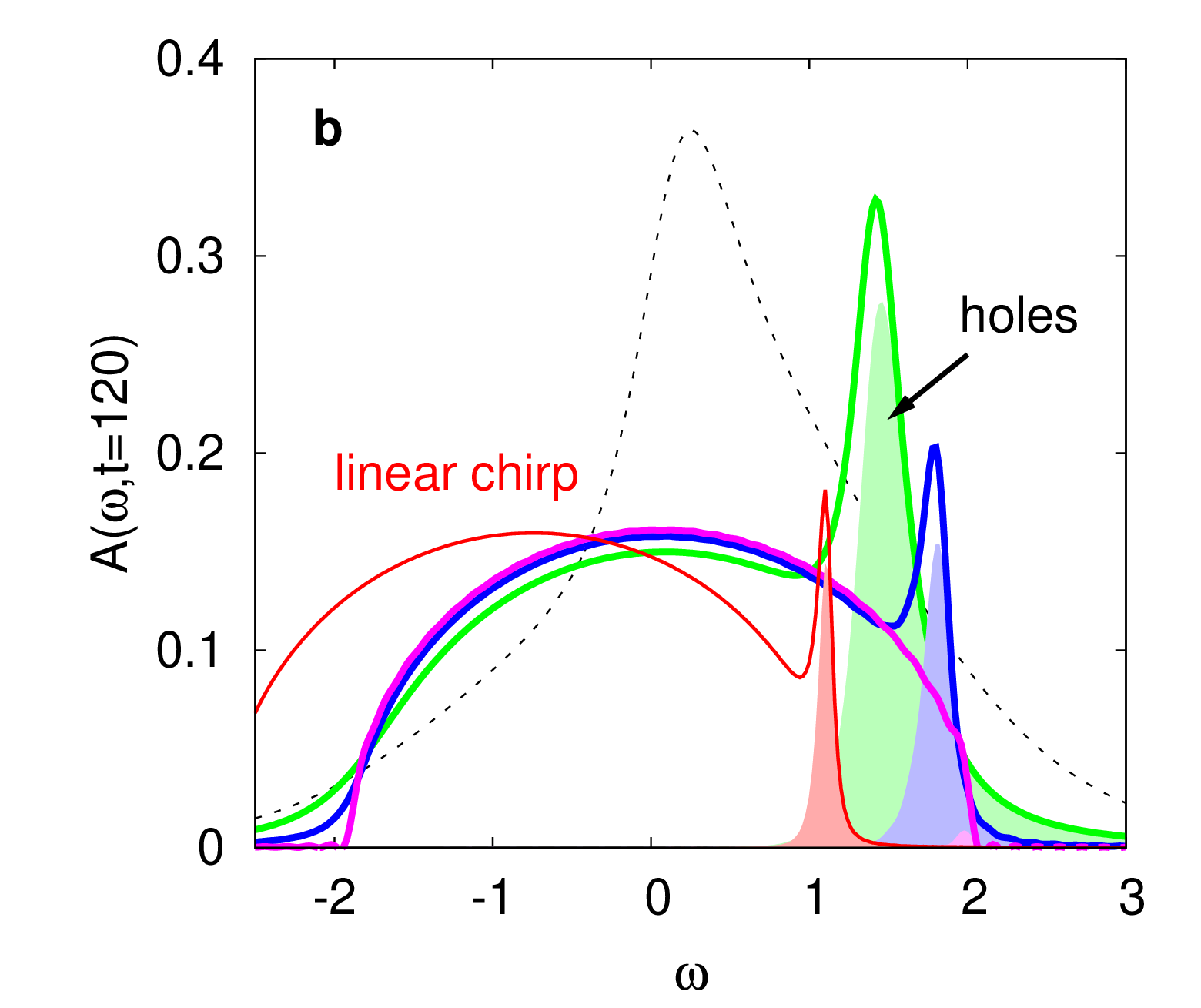}
\includegraphics[angle=-90, width=0.39\columnwidth]{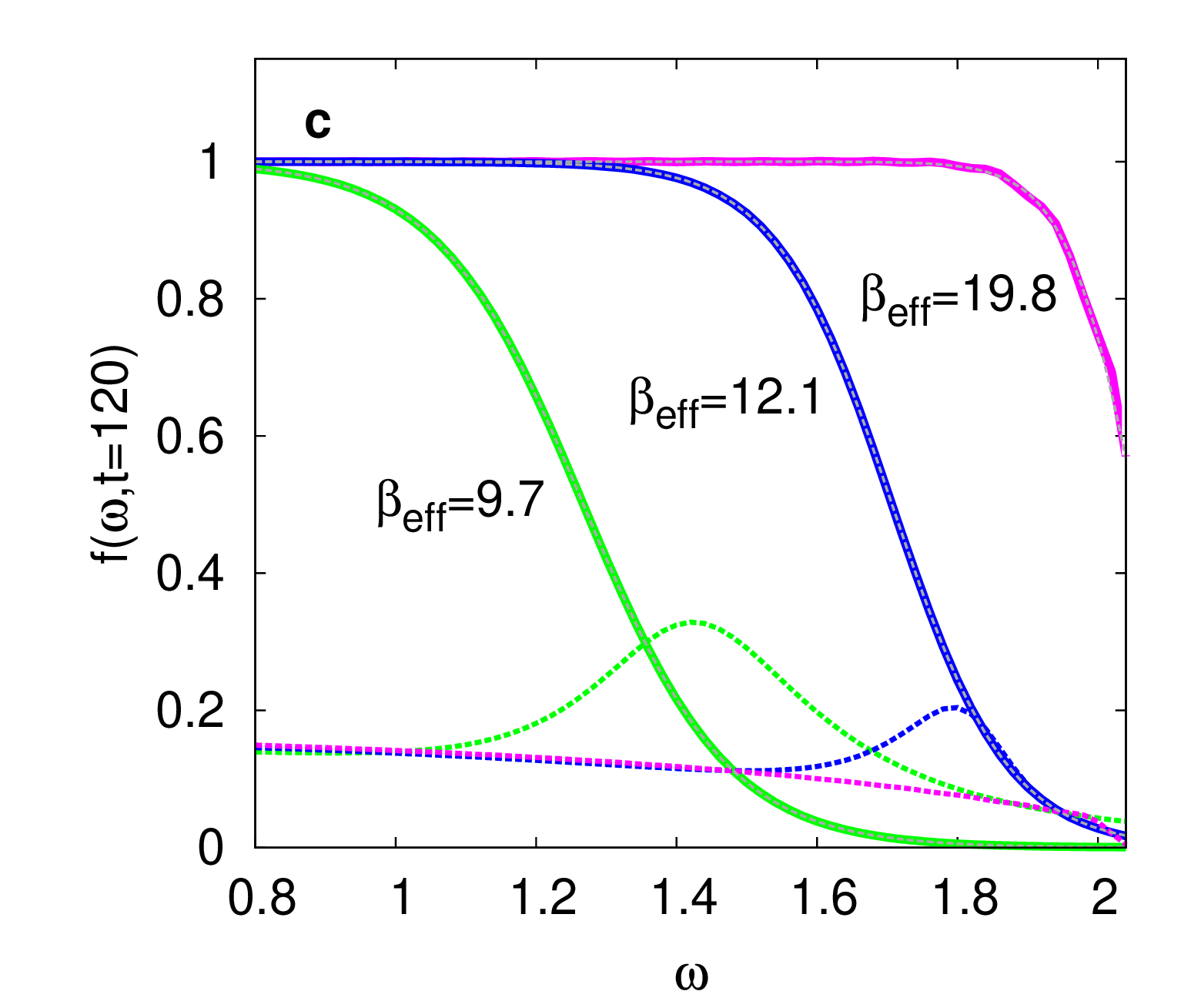}
\hspace{5mm}
\includegraphics[angle=-90, width=0.39\columnwidth]{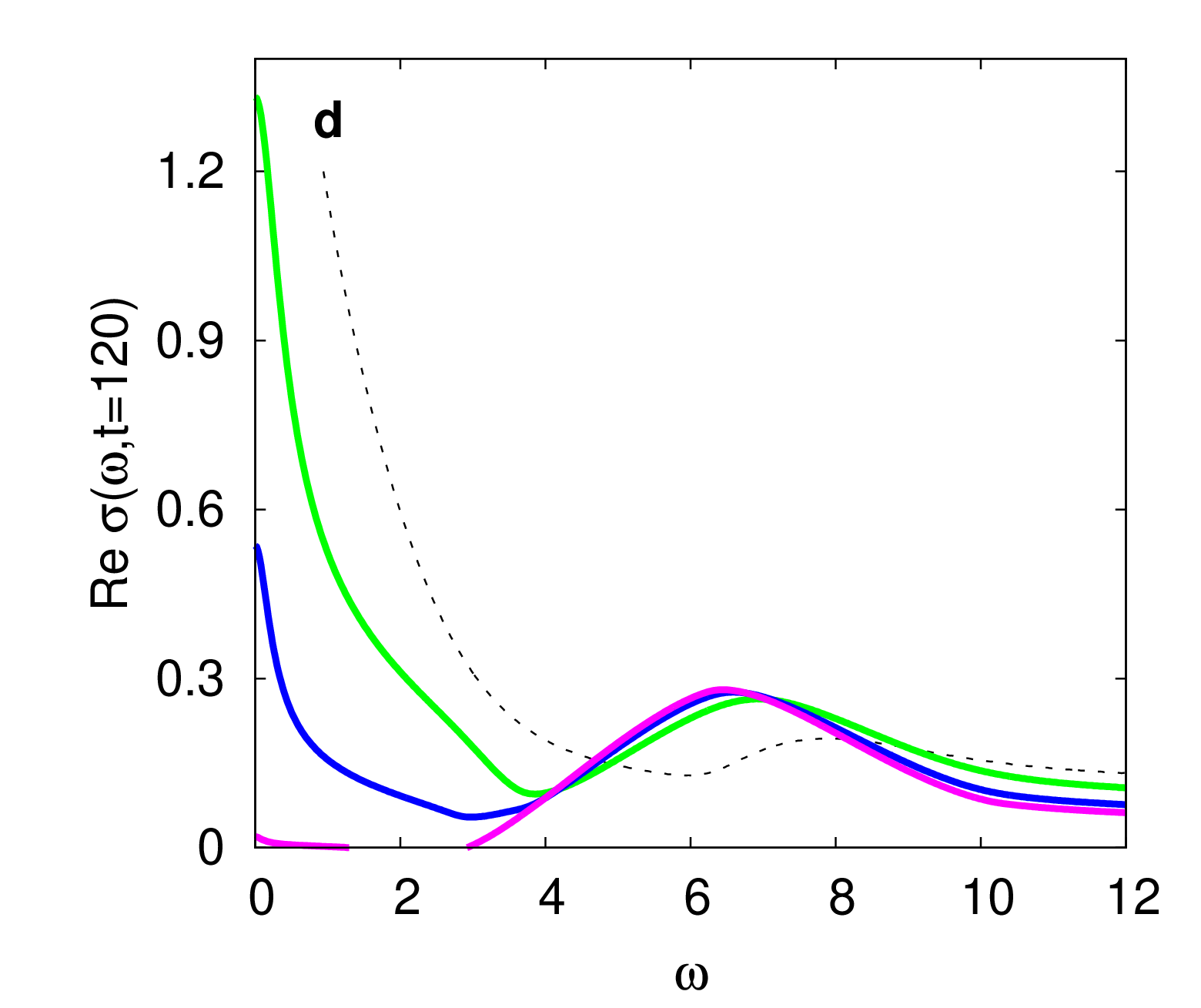}
\caption{
{\bf Charge transfer and quasi-particle cooling in the first set-up.} 
Panel {\bf a}: time evolution of the filling during a pulse with amplitude $a_\text{max}=0.8$ and indicated frequencies $\Omega$ and pulse durations $\delta$. 
Panel {\bf b}: Spectral functions (lower Hubbard band) measured after the longer pulse. Shaded regions show the hole occupation. 
Panel {\bf c}: Energy distribution function $f(\omega,t)$ measured after the longer pulse. The dashed gray lines are Fermi functions corresponding to the indicated inverse temperatures $\beta_\text{eff}$.  
Dashed colored lines show the spectral functions.
Panel {\bf d}: Real part of the optical conductivity measured after the longer pulse. The dashed black lines in panels b,d show the result for the initial quarter-filled state at $\beta=5$. The pulse amplitude is $a_\text{max}=0.8$. (The red spectrum in panel b shows the result for the second set-up and a chirped pulse of length $\delta=90$, amplitude $0.1625$, $\Omega_\text{in}=6.6$ and $\Omega_\text{fin}=7.4$.)  
}
\label{fig_pulse_dipole}
\label{fig_spectra}
\end{center}
\end{figure}  

The dipolar excitations in the chosen range of driving frequencies $\Omega$ lead to an increase in the occupation per spin $n_\sigma$ of the system. Figure~\ref{fig_pulse_dipole}a shows the  time evolution of $n_\sigma$ for different pulse frequencies and fixed pulse amplitude $a_\text{max}=0.8$. While the amplitude dependence is non-monotonous, longer pulses result in a larger increase of $n_\sigma$. The largest filling is obtained for $a_\text{max}=0.8$ and $\Omega\approx 6.8$. In this case, a pulse of duration $\delta=120$ yields an almost complete filling of the lower Hubbard band ($n_\sigma=0.499$), without any noteworthy increase in the number of doubly occupied sites. Hence, the initially quarter-filled metallic system is switched to an (almost) Mott insulating state via the photo-induced charge transfer from the core levels. This is the opposite effect from the usual photo-doping \cite{Eckstein2010photodoping, Eckstein2013,Iwai2003,Okamoto2010}, which transforms a Mott insulator into a nonthermal metal via charge excitations across the Mott gap. 

\begin{figure}[t]
\begin{center}
\includegraphics[angle=-90, width=0.39\columnwidth]{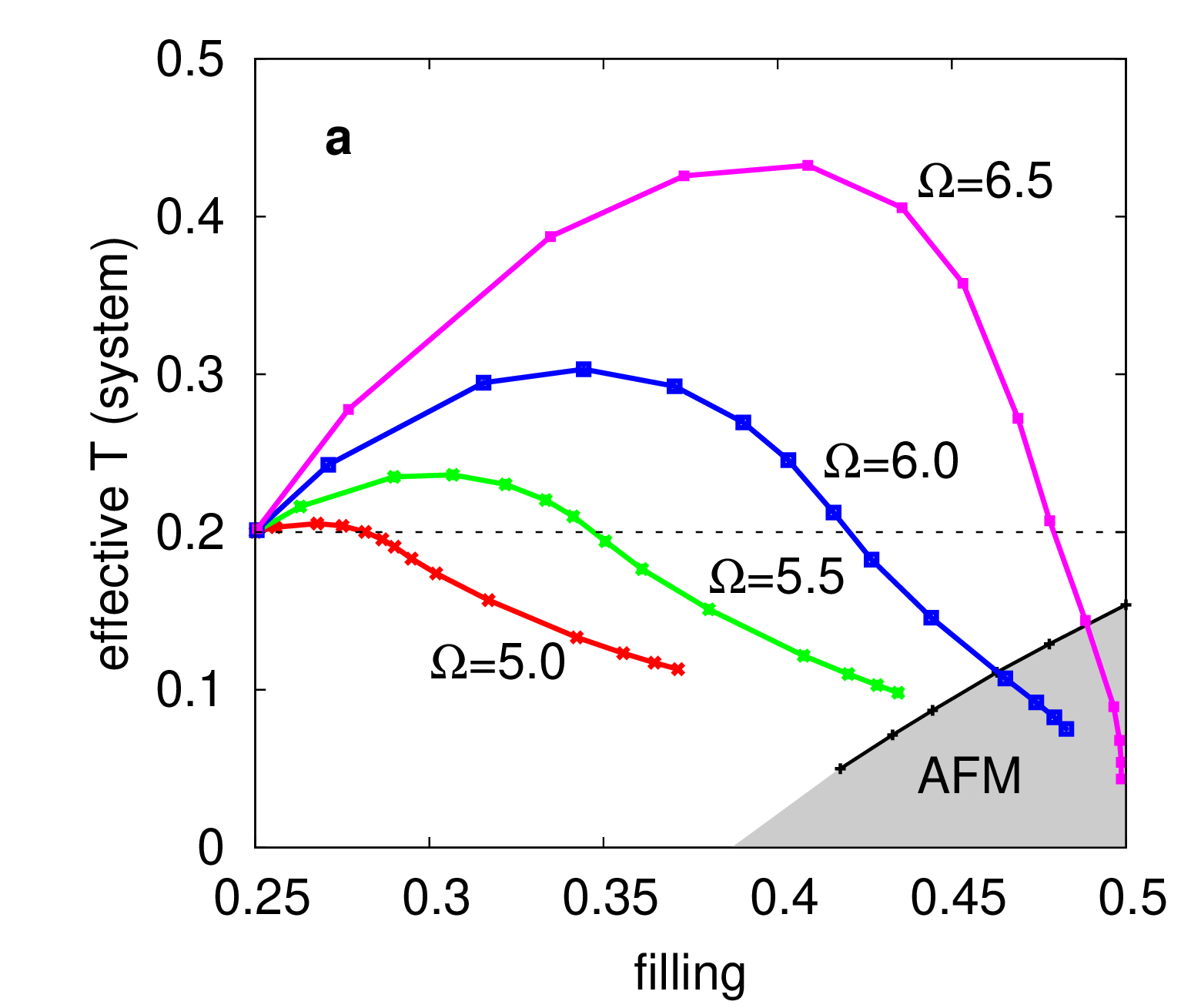}
\hspace{5mm}
\includegraphics[angle=-90, width=0.39\columnwidth]{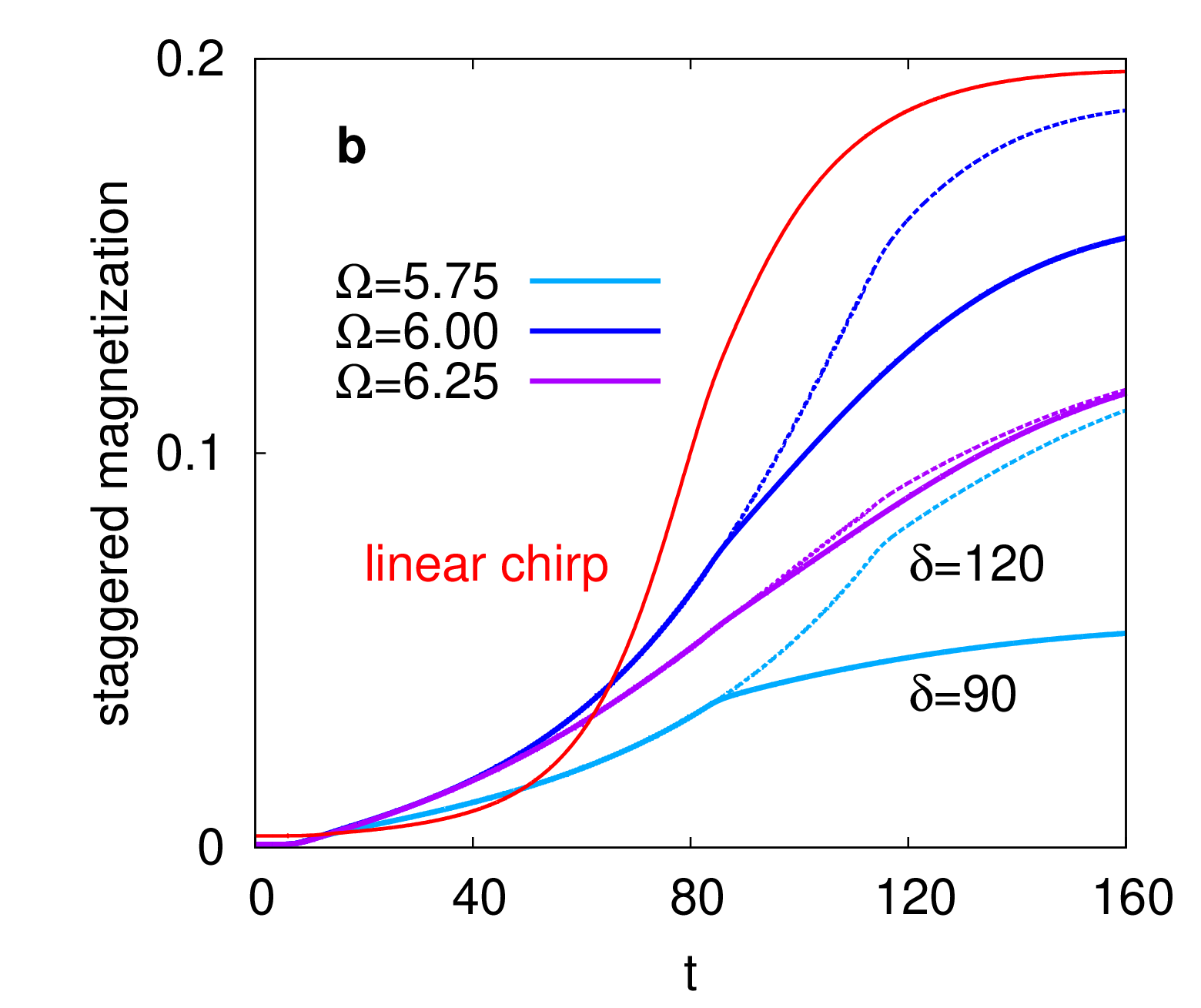}
\caption{{\bf Photo-doping induced antiferromagnetic order in the first set-up.} 
Panel {\bf a} plots the effective system temperature after the pulse as a function of filling. The dashed line shows the temperature ($T=0.2$) of the initial quarter-filled state. Different curves correspond to different (but fixed) pulse frequencies $\Omega$, while different points correspond to different pulse durations $\delta$ ($a_\text{max}=0.8$). The black line shows the N\'eel temperature in equilibrium.
Panel {\bf b}: Time dependence of the staggered magnetization for pulses with $\delta=90$ (solid lines) and 120 (dashed), $a_\text{max}=0.8$ and indicated $\Omega$ in the presence of a staggered magnetic field $h=0.001$. (The red curve in panel {\bf b} shows the result for the second set-up and a chirped pulse with parameters $\delta=90$, $a_\text{max}=0.1625$, $\Omega_\text{in}=6.6$ and $\Omega_\text{fin}=7.4$.)  
}
\label{fig_beta_eff}
\end{center}
\end{figure}

The light-induced metal-to-insulator transition is evident in the spectral function $A(\omega)$ and optical conductivity $\sigma(\omega)$, as illustrated in Figs.~\ref{fig_spectra}b,d, which show results measured immediately after the pulse of length $\delta=120$. The sharpness of the quasi-particle peaks in the spectral functions and of the Drude peaks in Re$\sigma(\omega)$ indicates a cold temperature of the photo-doped carriers. This is in stark contrast to the case of photo-doping across the Mott gap in a single-band Hubbard model, which typically results in hot charge carriers with a nonthermal distribution \cite{Eckstein2010photodoping} and very broad quasi-particle and Drude features \cite{Eckstein2013}. To demonstrate the thermal nature of the photo-doped system and extract the corresponding temperature $T_\text{eff}=1/\beta_\text{eff}$ and chemical potential $\mu_\text{eff}$, we plot in Fig.~\ref{fig_pulse_dipole}c the energy distribution functions $f(\omega,t)$ measured after different pulses with $\delta=120$ (see Methods). The dashed gray lines indicate fits to a Fermi distribution $f_F(\omega,\mu_\text{eff},\beta_\text{eff})=1/[1+\exp(\beta_\text{eff}(\omega-\mu_\text{eff}))]$, which very well match the measured distributions. We have confirmed that the equilibrium optical conductivities obtained for the measured $n_\sigma$ and $\beta_\text{eff}$ reproduce the results shown in Fig.~\ref{fig_spectra}d. Hence, the Hubbard subsystem thermalizes rapidly after the decoupling from the core levels.  

Remarkably, the effective temperature of the photo-doped Hubbard model can be substantially lower than that of the initial equilibrium state ($T=0.2$). In Fig.~\ref{fig_beta_eff}a we plot $T_\text{eff}$ measured for different pulse frequencies and pulse durations as a function of the filling after the pulse. Long pulses ($\delta \gtrsim 90$) with $\Omega=6.5$ and $a_\text{max}=0.8$ result in nearly half-filled systems with effective temperatures which are more than a factor of four lower than the initial temperature. $T_\text{eff}$ can drop below the N\'eel temperature, which is indicated by the ``AFM" line in Fig.~\ref{fig_beta_eff}a. Our results suggest that a state with antiferromagnetic long-range order can be realized in the present set-up by the combined effect of doping and cooling. 

To demonstrate the possibility of inducing long-range antiferromagnetic order we apply a small staggered magnetic field $h=0.001$ to the system. The time evolution of the staggered magnetization $\langle n_\uparrow-n_\downarrow\rangle$ is plotted for different pulse frequencies $\Omega$ and pulse durations $\delta=90$ and $120$ in Fig.~\ref{fig_beta_eff}b ($a_\text{max}=0.8$). Already during the photo-excitation, a symmetry-breaking is induced, which increases further after the end of the pulse. The magnetization however saturates at a lower value than the effective temperatures and fillings in Fig.~\ref{fig_beta_eff}a would suggest, which is due to the heating of the system during the symmetry-breaking process. Also, we notice a slow-down of the dynamics close to half-filling, due to the suppressed hopping. 

\begin{figure}[h!]
\begin{center}
\includegraphics[angle=-90, width=0.39\columnwidth]{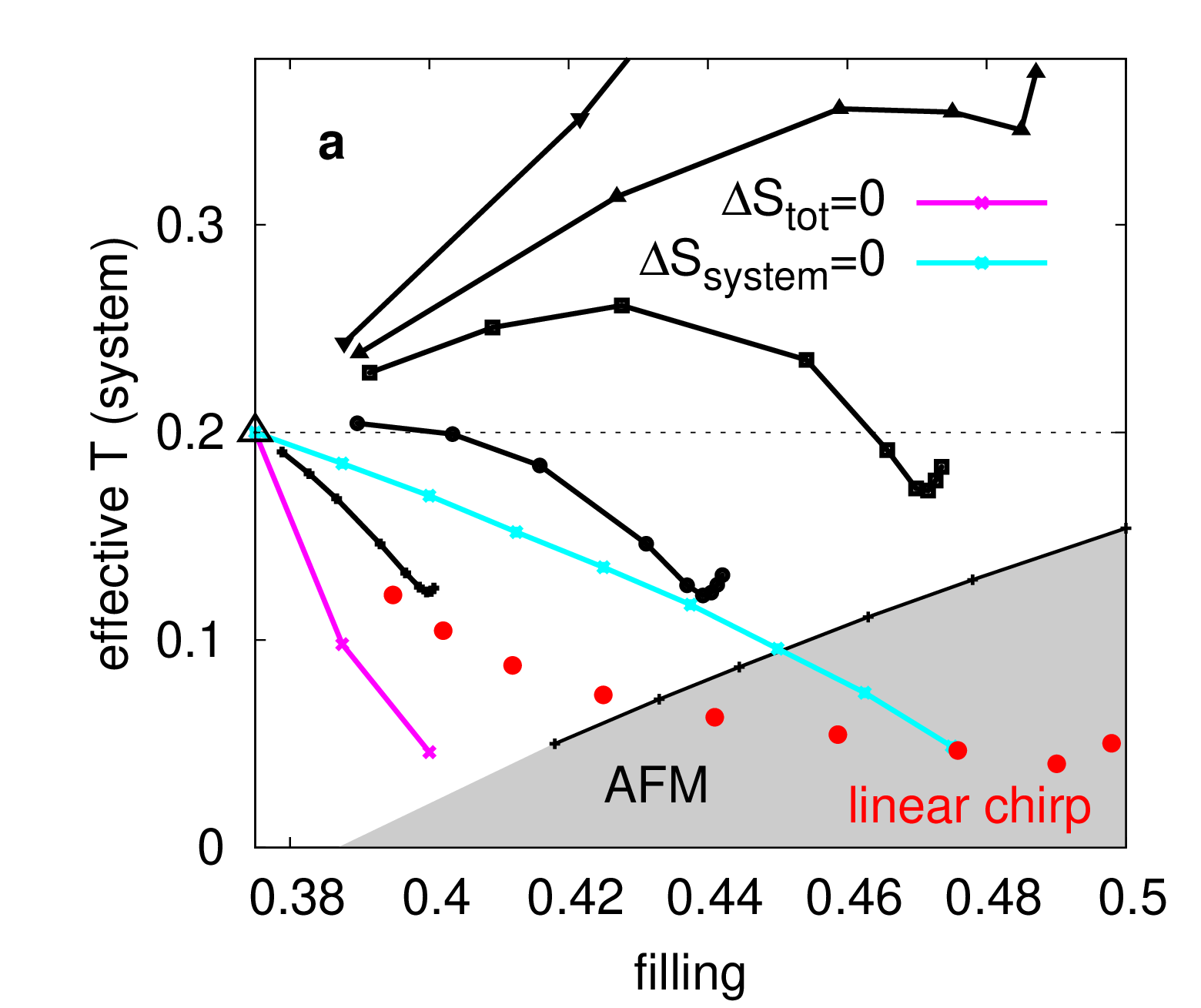}
\hspace{5mm}
\includegraphics[angle=-90, width=0.39\columnwidth]{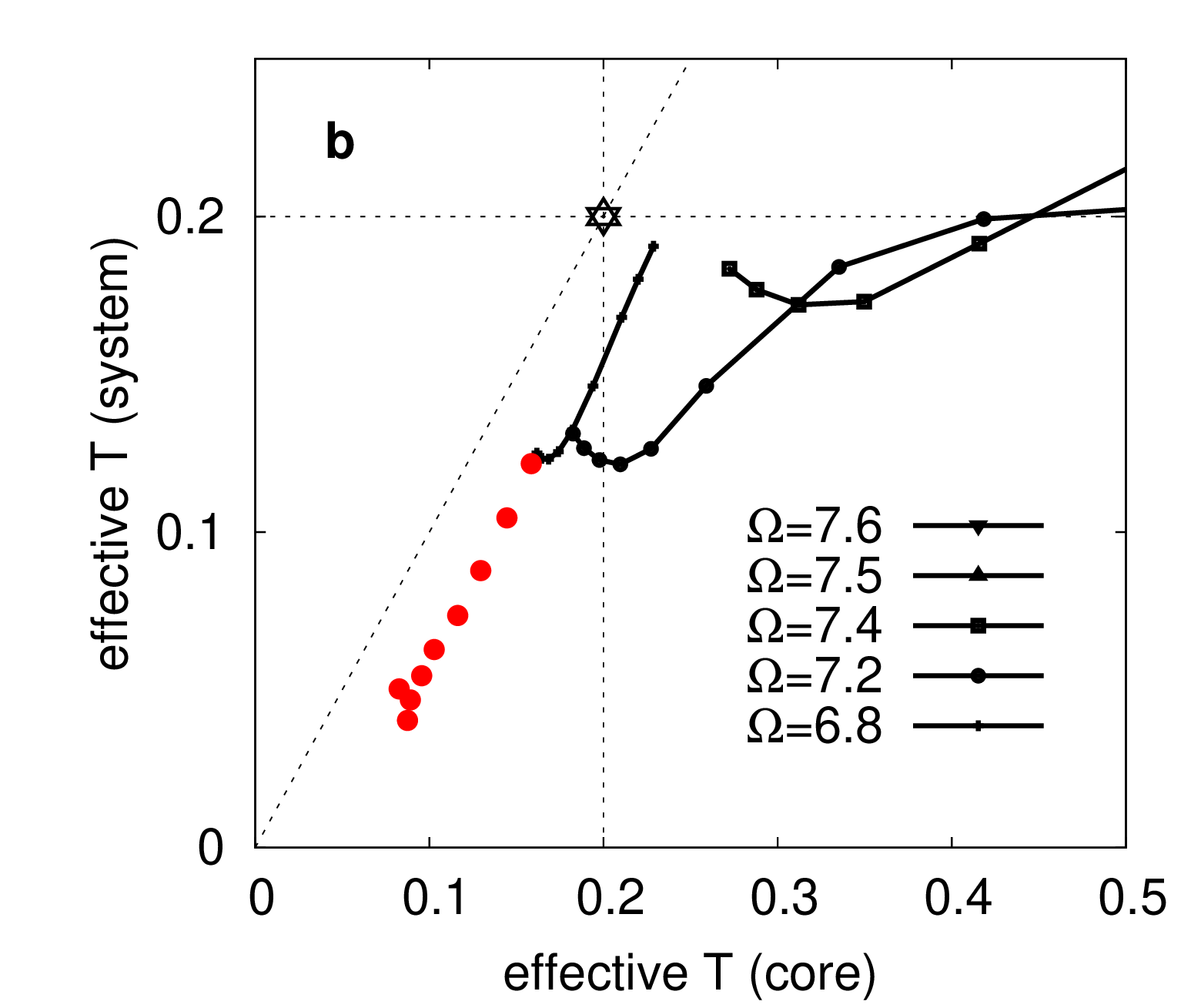}
\includegraphics[angle=-90, width=0.39\columnwidth]{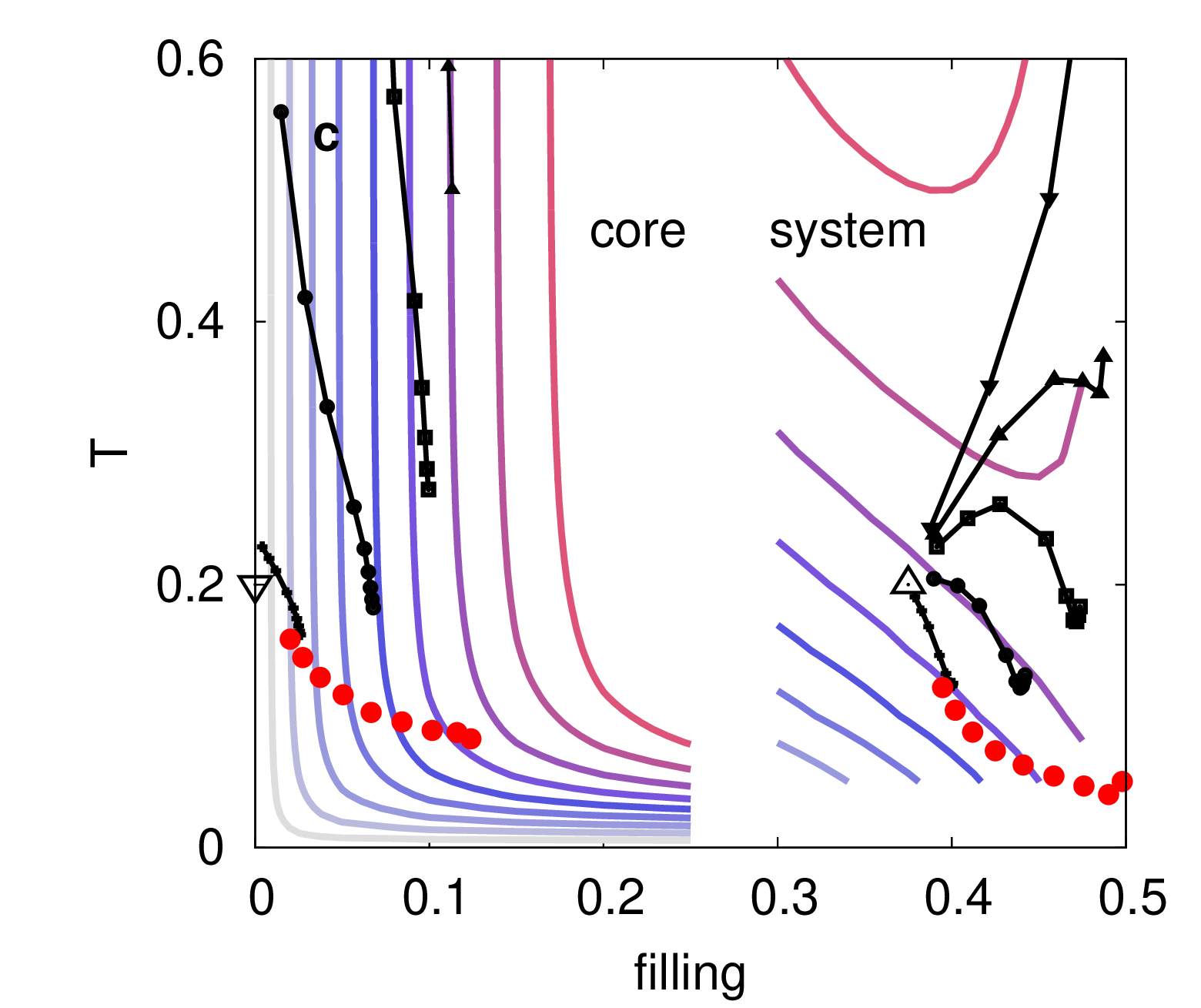}
\hspace{5mm}
\includegraphics[angle=-90, width=0.39\columnwidth]{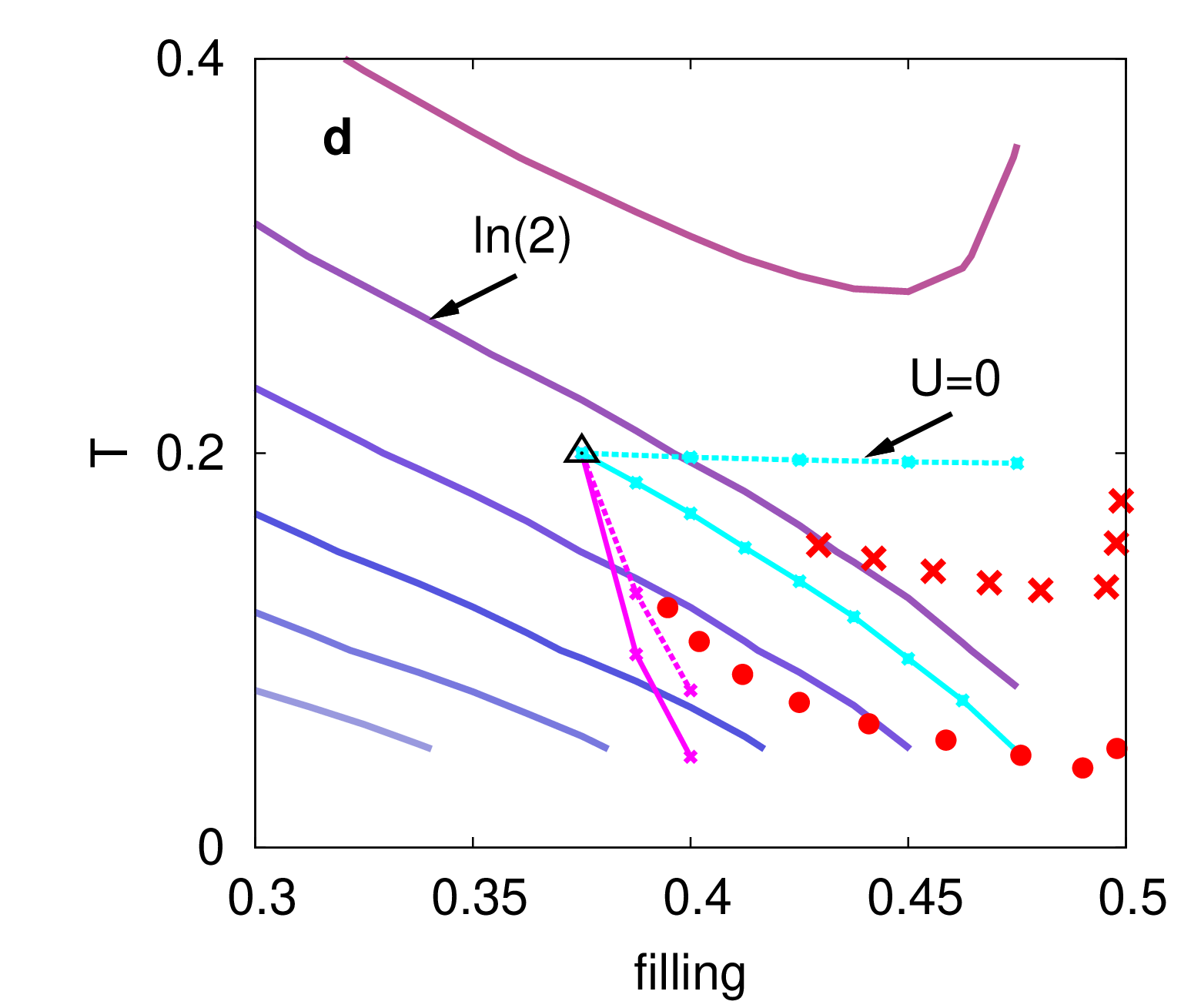}
\caption{
{\bf Cooling in the second set-up with a single narrow core band.}
Panel {\bf a}: Effective system temperature as a function of filling in a system with a single core band of width 0.4.  Different black symbols correspond to different driving frequencies $\Omega$, as indicated in panel {\bf b}, which plots the effective system temperature against the effective core band temperature. Different dots on a given black line correspond to different $a_\text{max}$.  Panel {\bf c} shows the isentropy curves in the space of filling and temperature. On the left, we show the result for the core band with $U=0$ and on the right for the Hubbard model with $U=6$. The contour lines correspond to $S=0.9$, $0.8$, \dots from top to bottom. Black curves indicate the core and system temperatures realized after photo-doping (for the core we show the hole density $1-n_\sigma$). 
Panel {\bf d}: Zoom of the low doping and low temperature region, with the $\Delta S_\text{tot}=0$ (magenta) and $\Delta S_\text{system}=0$ (cyan) curves indicated [solid (dashed) lines: $U_\text{system}=6$ ($0$)]. The red dots are the results obtained for a chirped pulse with $\Omega_\text{in}=6.6$, $6.7 \le \Omega_\text{fin} \le 7.4$ and optimized $a_\text{max}$. The red crosses in panel d show analogous results, but for a core bandwidth of 2, $\Omega_\text{in}=6.6$, $6.6 \le \Omega_\text{fin} \le 7.4$. All the data in this figure are for $\delta=90$.
}
\label{fig_entropy}
\end{center}
\end{figure}

Next, let us consider the second set-up in which the (noninteracting) and very narrow core band is equipped with a Bethe-lattice self-consistency (see Methods). To be able to reach fillings near $n_\sigma=0.5$, where the AFM order is most stable, we start from $n_\sigma=0.375$, as illustrated in Fig.~\ref{fig_illustration}b ($v_\text{system}=1$, $v_\text{core}=0.1$).  The black lines in Fig.~\ref{fig_entropy}a show $T_\text{eff}$ of the system after a pulse with $\delta=90$  and $\Omega=6.8, 7.2, 7.4, 7.5, 7.6$ (from left to right, see legend in panel b). Different dots on a given line correspond to different amplitudes in the range $0.025\le a_\text{max}\le 0.2$. The results demonstrate that for $\Omega \le 7.4$, the cooling-by-doping mechanism also works if we couple to a single core band, although the N\'eel temperature cannot be reached with $\delta=90$.  

The cooling mechanism can be understood by considering the entropy transfer between the two bands:  While the entropy in the core band is initially zero, it strongly increases with hole-doping. If the population transfer could be achieved at fixed {\em total} entropy, one could decrease the system entropy and temperature. In practice, it is hard to realize an isentropic population transfer, but we will now show that suitable protocols can nevertheless reduce the entropy of the system and that the corresponding reduction in temperature is enhanced by the fact that the system is driven towards the correlated Mott regime. 
For the quantitative analysis we compute the entropy per site $S$ of the noninteracting core band and of the Hubbard model as a function of filling and temperature $T$. Figure~\ref{fig_entropy}c  shows the contour lines of the entropy for the core band, plotted against hole doping, and for the interacting system ($U=6$), plotted against the electron occupancy. The results for $U=0$ have been obtained numerically by integrating $C/T$ ($C$ is the specific heat per site) from $T=0$ and those for $U=6$ by integrating $C/T$ down from the high-$T$ limit. The isentropic lines of the core band are almost vertical for temperatures above a scale set by its narrow bandwidth, corresponding to the infinite temperature result $S_{\infty}=-2n_\sigma\ln(n_\sigma)-2(1-n_\sigma)\ln(1-n_\sigma)$. In the regime of interest to our simulation, the entropy increase of the core band is therefore approximately given by $S_{\infty}$. This is confirmed by the black and red symbols, which show the effective temperatures and hole densities reached by different photo-doping protocols. The entropy of the valence band is more interesting (see right part of Fig.~\ref{fig_entropy}c). At low temperature, in a Fermi liquid, we expect $S(T)=\gamma T$ with $\gamma=\lim_{T\rightarrow 0}C/T$. In the doped Mott regime, the $\gamma$-factor diverges like $\gamma\propto 1/|0.5-n_\sigma|$ near half-filling \cite{Werner2007}, while the Fermi liquid coherence temperature (and hence the $T$-range for which $S_\text{system}(T)=\gamma T$ holds) drops. In the paramagnetic insulator at $n_\sigma=0.5$, the entropy is roughly given by the remaining spin entropy $S=\ln(2)$ down to the lowest temperature, while states contributing to entropy $S>\ln(2)$ are only activated by charge fluctuations at  temperatures of the order of $U$. This explains why the isentropic curves  for $S\lesssim\ln (2)$ decrease towards zero near $n_\sigma=0.5$ while they increase with density for $S\gtrsim\ln (2)$, leading to a non-monotonous behavior of the isentropy curves as a function of filling for intermediate entropies. 

An isentropic photo-doping process, in which the increase $\Delta S_\text{core}\approx \Delta S_{\infty}$ of the core entropy is compensated by a corresponding decrease of the system entropy, leads to a drastic reduction of the temperature as a function of the transferred charge (see solid magenta line in Fig.~\ref{fig_entropy}a,d). While a realistic process will never conserve the total entropy, the slope of the isentropy lines close to the Mott regime implies that even if the system entropy remained constant or increased slightly during a photo-doping process starting from the Fermi liquid regime, the system temperature can still decrease (see solid cyan curve  in Fig.~\ref{fig_entropy}a,d, which shows the line corresponding to $\Delta S_\text{system}=0$).  As indicated by the solid black lines, for low enough $\Omega$, there is indeed a decrease of entropy in the system, i.e., a reshuffling of entropy from the system to the core band, and the cooling of the system can be much stronger than in the $\Delta S_\text{system}=0$ case.  

If the system were noninteracting ($U=0$), the isentropy lines of the system would be almost horizontal on the scale of Fig.~\ref{fig_entropy}d, see dashed cyan curve, suggesting little or no cooling under realistic driving conditions. 
In the idealized adiabatic case, $T_\text{eff}$ would be controlled by the density of states $\rho$ at the (filling-dependent) Fermi energy, since $T_\text{eff}(\rho_\text{system}+\rho_\text{core})$ is constant. 
The cooling effect would still be strong (dashed magenta line), but less pronounced than for the interacting system. The isentropy curves in panels c,d also explain why the photo-doping of the cold, half-filled Mott insulator generically results in strong heating: Here one starts with a large entropy of $\ln (2)$ per site, which is much larger than the entropy of a cold, Fermi-liquid-like metal. Hence, assuming that the entropy of a photo-doped insulator is similar to that of a chemically doped insulator, we conclude that even in the isentropic case, $T_\text{eff}$ of the photo-doped state must strongly increase. 

Photo-excitation with low $\Omega$ creates steep (and hence cold) distribution functions $f(\omega,t)$ by filling high-kinetic energy holes in the correlated system with electrons from the core band, analogous to evaporative cooling, but the amount of transferred charge is limited. In order to create a strongly correlated Fermi liquid with low temperature, we apply a {\it chirped} pulse of the form $a(t)=a_\text{max}\sin[\Omega(t) t]$ with $\Omega(t)$ increasing linearly from $\Omega_\text{in}$ to $\Omega_\text{fin}$ during the pulse of duration $\delta$. The result for $\Omega_\text{in}=6.6$, $\Omega_\text{fin}=0.67$, $0.68$,\ldots, $7.4$ and $\delta=90$ is shown by the red dots in Fig.~\ref{fig_entropy}. Here we have adjusted the amplitude $a_\text{max}$ to reach the lowest $T_\text{eff}$ of the system for each $\Omega_\text{fin}$. (The lowest temperature of $T_\text{eff}=0.04$ is reached for $\Omega_\text{fin}=7.4$ and $a_\text{max}=0.1625$; the gradual filling of the band with this type of pulse is illustrated in Fig.~\ref{fig_illustration}d.) With chirped pulses, we can easily reach the strongly correlated Fermi liquid regime  and trigger a symmetry breaking to the AFM phase. The spectral function obtained after the $\Omega_\text{fin}=7.4$, $a_\text{max}=0.1625$ pulse is shown by the red curve in Fig.~\ref{fig_spectra}b, and the magnetization dynamics in the presence of a seed field $h=0.001$ by the red curve in Fig.~\ref{fig_beta_eff}b. 
The width of the core band matters because it sets the temperature scale where the isentropic curves of the core bend from vertical to horizontal (see Fig.~\ref{fig_entropy}c). For a wider core bandwidth of 2, the bending of the blue isentropy curve occurs around $T\approx 0.4$ and as a result, the increase of the core entropy with hole doping at fixed $T=0.2$ is reduced. Indeed, for the same chirping protocol, the model with the wider core band produces a substantially smaller cooling effect (see red crosses in Fig.~\ref{fig_entropy}d). 
We also note that both the system and core can be cooled (Fig.~\ref{fig_entropy}b), and that in the chirped case their effective temperatures are similar. Interband scattering processes, or weaker driving, would result in an even better synchronization. 

Our simulations show that the charge transfer induced by dipolar excitations between a filled core band and a partially filled Hubbard band can be accomplished in such a way that entropy is shifted to the core band. It is interesting to compare this evaporative hole cooling mechanism to the established technique of thermal spin mixing \cite{Abragam1978}, used to efficiently cool electronic spins. There, a narrow band  is formed by spin excitations that propagate via dipolar spin flip-flop terms \cite{Rodriguez2018, DeLuca2015, DeLuca2016}.  By driving such excitations off the band center, and letting them relax with a phonon-bath, a steady state with an effective temperature of the order of the spin bandwidth is created. The evaporative cooling instead achieves low temperatures by using the narrow band as an efficient entropy sink. It could be interesting to explore combinations of the two mechanisms to optimize cooling protocols.

The set-ups explored in the present study provide a means to realize complex phases such as correlated Fermi liquids, antiferromagnetic, superconducting or excitonic order by the combined effect of doping and cooling. The basic strategy is applicable to cold atom and condensed matter systems. In cold atom experiments, the two subsystems could be realized in separate layers, which are transiently coupled by a periodic hopping modulation or a shift in the local energies and tunnel barriers. Since the threshold entropy for antiferromagnetic order has recently been reached\cite{Mazurenko2017}, the additional cooling provided by the particle transfer could give access to the sought-after pseudo-gap and superconducting phases. In correlated materials, it is important to identify suitable (full or empty) flat bands, and to exploit the asymmetry in the laser excitations into and out of the bands to realize the reshuffling of entropy.  

\mbox{}\\
\noindent
{\bf Methods}

We use nonequilibrium DMFT\cite{Aoki2014} to simulate the charge transfer from the core band to the system via dipolar excitations. The system is described by the Hubbard model $H_\text{system}(t)=\tilde v(t)\sum_{\langle i,j\rangle \sigma} (c^\dagger_{i\sigma} c_{j\sigma} + \text{h.c.})+U\sum_{i}n_{i\uparrow}n_{i\downarrow}-\mu\sum_i(n_{i\uparrow}+n_{i\downarrow})$ on an infinitely connected Bethe lattice. Here, $c^\dagger_{i\sigma}$ creates a fermion on site $i$ with spin $\sigma$, $U$ is the Hubbard repulsion, $\tilde v$ the hopping between nearest neighbor sites, and $\mu$ the chemical potential. In DMFT\cite{Georges1996} this lattice model is mapped onto a quantum impurity model with action $S_\text{imp}[U,\Delta_\text{system}]$ defined by the on-site interaction $U$ and a hybridization function $\Delta_\text{system}$. The latter is determined self-consistently in such a way that the Green function of the impurity model reproduces the local Green function of the lattice model\cite{Georges1996,Aoki2014}: $G_\text{system}(t,t')\equiv -i\text{Tr}[T_\mathcal{C}e^{S_\text{imp}}c(t)c^\dagger(t')]/\text{Tr}[T_\mathcal{C}e^{S_\text{imp}}]$. To evaluate the right hand side for a given $\Delta_\text{system}$, we use the non-crossing approximation (NCA)\cite{Keiter1971,Eckstein2010}. The DMFT solution (with an exact impurity solver) becomes exact in the limit of infinite coordination number $z$, if the hopping is rescaled as $v_\text{system}=\tilde v/\sqrt{z}$. In the case of the Bethe lattice the self-consistency condition for the model without core band simplifies to $\Delta_{\text{system},\sigma}(t,t')=v_\text{system}(t)G_{\text{system},\sigma}(t,t')v_\text{system}(t')$.

In the first set-up, the effect of the core band is described by an additional hybridization function $\Delta_\text{core}(t,t')=v_\text{system-core}(t)G^0_\text{core}(t,t')v^*_\text{system-core}(t')$, where $G^0_{\text{core}}$ is the Green function associated with the DOS of the core band. Since the core band remains in equilibrium (i.e. fully filled), there is only one self-consistency equation for the hybridization function of the system:
\begin{equation}
\Delta_{\text{system},\sigma}(t,t')=v_\text{system}(t) G_{\text{system},\sigma}(t,t')v^*_\text{system}(t')+v_\text{system-core}(t)G^0_\text{core}(t,t')v_\text{system-core}^*(t').
\end{equation}
Dipolar excitations from the core to the system are described by a time-periodic modulation $v_\text{system-core}(t)=a(t)f(t-t_p)$, where $a(t)=a_\text{max}\sin(\Omega t)$ is an oscillating function with frequency $\Omega$ and amplitude $a_\text{max}$ and $f(t)=1/[(1+\exp(t-\delta+6))(1+\exp(-t+6))]$ a pulse envelope function of width $\delta$, with a ramp on and off time of approximately $2\times 6$. For $\Omega$ comparable to the energy difference between the core levels and the lower Hubbard band, this photo-excitation results in a transfer of charge from the core levels to the system. We describe the core levels by a box-shaped DOS with smooth edges in the energy range $-7\le \omega \le -5$ and consider driving frequencies $5\le \Omega\le 7$ and pulses of duration $\delta \approx 30-120$. These multi-cycle pulses produce only a negligible number of double occupations, i.e. the upper Hubbard band remains empty.  

The optical conductivity $\sigma$ of the system can be probed by dressing $v_\text{system}(t)$ with an appropriate Peierls factor\cite{Aoki2014} (For the implementation of the electric field in the Bethe lattice see Ref.~\onlinecite{Werner2017}.) We apply a short electric field pulse $E_t(t')$ centered at a time $t=\delta$, i.e. immediately after the photo-doping pulse, and measure the induced current $j_t(t')$.  After Fourier transformation, one obtains $\sigma(\omega,t)=j_t(\omega)/E_t(\omega)$. The time-dependent spectral function $A(\omega,t)$ is calculated by forward integration of the retarded Green function, $A(\omega,t)=-\frac{1}{\pi}\text{Im}\int_t^{t_\text{max}} dt'e^{i\omega (t'-t)}G^\text{ret}(t',t)$, and similarly the time-dependent occupation function $A^<(\omega,t)$ is calculated from the lesser component~\cite{Aoki2014}. The ratio $f(\omega,t)=A^<(\omega,t)/A^\text{ret}(\omega,t)$ defines the time-dependent distribution function, which in a thermalized state becomes a Fermi distribution function. AFM order can be studied in DMFT by considering two sublattices with opposite spin polarization and flipping the spin index of the hybridization function in the self-consistency equation\cite{Georges1996}. 

In the second set-up, we describe the core band by a noninteracting Hubbard model ($U=0$) of bandwidth 0.4, which like the system is solved with DMFT+NCA (to enable equilibration of the core electrons). 
In this case, the core band adds a hybridization term to system's impurity model and vice versa. The two coupled DMFT self-consistency equations become 
\begin{align}
\Delta_{\text{system},\sigma}(t,t')&=v_\text{system}(t) G_{\text{system},\sigma}(t,t') v_\text{system}(t')+v_\text{system-core}(t)G_{\text{core},\sigma}(t,t')v_\text{system-core}^*(t'),\\
 \Delta_{\text{core},\sigma}(t,t')&=v_\text{core}(t) G_{\text{core},\sigma}(t,t') v_\text{core}(t')+v_\text{system-core}(t)G_{\text{system},\sigma}(t,t')v_\text{system-core}^*(t').
\end{align}

The effect of the periodic driving of $v_\text{system-core}$ amounts to a shift of the core level DOS by $\Omega$ (see dashed lines in Fig.~\ref{fig_illustration}a,c). We have also performed simulations in which the core levels (and their chemical potential) are shifted by the energy $\epsilon=\Omega$  and the hopping between the core levels and the system is smoothly switched on and off. The results are qualitatively similar to those reported in the main text. 

\clearpage

{\it Acknowledgements} We thank A. J. Millis for helpful discussions. The calculations have been performed on the Beo04 cluster at the University of Fribourg, using a software library co-developed by H. Strand. This work has been supported by the European Research Council through ERC Consolidator Grant No.~724103 (PW)
and ERC Starting Grant No.~716648 (ME). GR acknowledges the support from the ARO MURI W911NF-16-1-0361 Quantum Materials by Design with Electromagnetic Excitation sponsored by the U.S. Army, from the Institute of Quantum Information and Matter, an NSF Frontier center funded by the Gordon and Betty Moore Foundation, and from the Packard Foundation. This work was initiated at the Aspen Center for Physics, during the 2018 Summer Program. PW and GR thank the ACP for its hospitality. 




\begin{thebibliography}{99}
 
\newcommand{\mytitle}[1]{``#1'',}

%
%
%
%
%
%
%
%
%
%
%
%
%
%
%
%
%
%
%
%


\bibitem{Denny2015} Denny, S. J.,  Clark, S. R., Laplace, Y., Cavalleri, A. \& Jaksch, D. Proposed Parametric Cooling of Bilayer Cuprate Superconductors by Terahertz Excitation. {\it Phys. Rev. Lett} {\bf 114}, 137001 (2015). 

\bibitem{Okamoto2016} Okamoto, J., Cavalleri, A. \& Mathey, L. Theory of Enhanced Interlayer Tunneling in Optically Driven High-Tc Superconductors {\it Phys. Rev. Lett.} {\bf 117}, 227001 (2016).

\bibitem{Sentef2017} Sentef, M. A., Tokuno, A., Georges, A. \& Kollath, C. Theory of laser-controlled competing superconducting and charge orders. {\it Phys. Rev. Lett.} {\bf 118}, 087002 (2017).

\bibitem{Murakami2017} Murakami, Y., Tsuji, N., Eckstein, M. \& Werner, P. Nonequilibrium steady states and transient dynamics of conventional superconductors under phonon driving. {\it Phys. Rev. B} {\bf 96}, 045125 (2017). 

\bibitem{Babadi2017} Babadi, M., Knap, M., Martin, I., Refael, G \& Demler, E. Theory of parametrically amplified electron-phonon superconductivity, {\it Phys. Rev. B} {\bf 96}, 014512 (2017).

\bibitem{Claassen2017} Claassen, M., Jiang, H.-C., Moritz, B. \& Devereaux, T. P. Dynamical time-reversal symmetry breaking and photo-induced chiral spin liquids in frustrated Mott insulators. {\it Nat. Commun.} {\bf 8}, 1192 (2017).

\bibitem{Kennes2017} Kennes, D. M., Wilner, E. Y., Reichman, D. R., \& Millis, A. J. Transient superconductivity from electronic squeezing of optically pumped phonons. {\it Nat. Phys.} {\bf 13}, 479 (2017).

\bibitem{Murakami2017exc} Murakami, Y., Golez, D., Eckstein, M. \& Werner, P. Photoinduced enhancement of excitonic order. {\it Phys. Rev. Lett.} {\bf 119}, 247601 (2017).

\bibitem{Mazza2017} Mazza, G. \& Georges, A. Non-equilibrium superconductivity in driven alkali-doped fullerides. {\it Phys. Rev. B} {\bf 96}, 064515 (2017).

\bibitem{Nava2018} Nava, A., Giannetti, G., Georges, A., Tosatti, E., \& Fabrizio, M., Cooling quasiparticles in A$_3$C$_{60}$ fullerides by excitonic mid-infrared absorption. {\it Nature Physics} {\bf 14}, 154 (2018).

\bibitem{Fabrizio2018} Fabrizio, M. Selective Transient Cooling by Impulse Perturbations in a Simple Toy Model. {\it Phys. Rev. Lett.} {\bf 120}, 220601 (2018). 

\bibitem{Li2018} Li, J., Strand, H., Werner, P. \& Eckstein, M. Theory of photoinduced ultrafast switching to a spin-orbital ordered `hidden' phase. {\it Nature Commun.} {\bf 9}, 4581 (2018).

\bibitem{Kaiser2014} Kaiser, S., Hunt, C. R., Nicoletti, D., Hu, W., Gierz, I., Liu, H. Y., Le Tacon, M., Loew, T., Haug, D., Keimer, B. \& Cavalleri, A. Optically induced coherent transport far above Tc in underdoped YBa$_2$Cu$_3$O$_{6+\delta}$. {\it Phys. Rev. B} {\bf 89}, 184516 (2014).

\bibitem{Hu2014} Hu, W., Kaiser, S., Nicoletti, D., Hunt, C. R., Gierz, I., Hoffmann, M. C., Le Tacon, M., Loew, T., Keimer, B., \& Cavalleri, A. Optically enhanced coherent transport in YBa$_2$Cu$_3$O$_{6.5}$ by ultrafast redistribution of interlayer coupling. {\it Nat. Mater.} {\bf 13}, 705 (2014).

\bibitem{Mitrano2016} Mitrano, M., Cantaluppi, A., Nicoletti, D., Kaiser, S., Perucchi, A., Lupi, S., Di Pietro, P., Pontiroli, D.,  Ricco, M., Clark, S. R., Jaksch, D., \& Cavalleri, A. Possible light-induced superconductivity in K$_3$C$_{60}$ at high temperature. {\it Nature} {\bf 530}, 461 (2016).   

\bibitem{Aoki2014} Aoki, H., Tsuji, N., Eckstein, M., Kollar, M., Oka, T. \& Werner, P. Nonequilibrium dynamical mean-field theory and its applications. {\it Rev. Mod. Phys.} {\bf 86}, 779 (2014).
  
\bibitem{Herrmann2017} Herrmann, A., Murakami, Y., Eckstein, M., \& Werner, P. Floquet prethermalization in the resonantly driven Hubbard model. Europhys. Lett. {\bf 120}, 57001 (2018).

\bibitem{Werner2012} Werner, P., Tsuji, N., \& Eckstein, M. Nonthermal symmetry-broken states in the strongly interacting Hubbard model. {\it Phys. Rev. B} {\bf 86}, 205101 (2012).

\bibitem{Golez2016} Golez, D., Werner, P. \& Eckstein, M. Photo-induced gap closure in an excitonic insulator. Phys. Rev. B {\bf 94}, 035121 (2016).

\bibitem{Eckstein2013} Eckstein M. \& Werner, P. Photoinduced States in a Mott Insulator. {\it Phys. Rev. Lett.} {\bf 110}, 126401 (2013).

\bibitem{Werner2018} Werner, P., Strand, H., Hoshino, S., Murakami, Y. \& Eckstein, M. Enhanced pairing susceptibility in a photodoped two-orbital Hubbard model. {\it Phys. Rev. B} {\bf 97}, 165119 (2018).


\bibitem{Bernier2009} Bernier, J.-S., Kollath, C., Georges, A., De Leo, L., Gerbier, F., Salomon, C. \& K\"ohl, C. Cooling fermionic atoms ins optical lattices by shaping the confinement. {\it Phys. Rev. A} {\bf 79}, 061601(R) (2009). 

\bibitem{Chiu2018} Chiu, C. S., Ji, G., Mazurenko, A., Greif, D., \& Greiner, M. Quantum State Engineering of a Hubbard System with Ultracold Atoms. {\it Phys. Rev. Lett.} {\bf 120}, 243201 (2018). 

\bibitem{Georges1996} Georges, A., Kotliar, G., Krauth, W. \& Rozenberg, Dynamical mean-field theory of strongly correlated fermion systems and the limit of infinite dimensions. M. J. {\it Rev. Mod. Phys.} {\bf 68}, 13 (1996).

\bibitem{Keiter1971} Keiter, H. \& Kimball, J. C. Diagrammatic Approach to the Anderson Model for Dilute Alloys. {\it Journal of Applied Physics} {\bf 42}, 1460 (1971)

\bibitem{Eckstein2010} Eckstein, M. \& Werner, P. Nonequilibrium dynamical mean-field calculations based on the noncrossing approximation and its generalizations. {\it Phys. Rev. B} {\bf 82}, 115115 (2010).

\bibitem{Werner2017} Werner, P., Hoshino, S., Strand, H. \& Eckstein, M., Ultrafast switching of composite order in A$_3$C$_{60}$, {\it Phys. Rev. B} {\bf 95}, 195405 (2017).

\bibitem{Eckstein2010photodoping} Eckstein, M. and Werner, P. Thermalization of a pump-excited Mott insulator. {\it Phys. Rev. B} {\bf 84}, 035122 (2010).

\bibitem{Iwai2003} Iwai, S., Ono, M., Maeda, A., Matsuzaki, H., Kishida, H., Okamoto, H., and Tokura, Y. Ultrafast Optical Switching to a Metallic State by Photoinduced Mott Transition in a Halogen-Bridged Nickel-Chain Compound. {\it Phys. Rev. Lett.} {\bf 91}, 057401 (2003).

\bibitem{Okamoto2010} Okamoto, H., Miyagoe, T., Kobayashi, K., Uemura, H., Nishioka, H., Matsuzaki, H., Sawa, A. \&  Tokura, Y. Ultrafast charge dynamics in photoexcited Nd$_2$CuO$_4$ and La$_2$CuO$_4$ cuprate compounds investigated by femtosecond absorption spectroscopy. {\it Phys. Rev. B} {\bf 82}, 060513 (2010).

\bibitem{Werner2007} Werner, P. \& Millis, A. J. Doping-driven Mott transition in the one-band Hubbard model. {\it Phys. Rev. B} {\bf 75}, 085108 (2007).

\bibitem{Abragam1978} Abragam, A. \& Goldman, M. Principles of dynamic nuclear polarisation. {\it Rep. Prog. Phys.} {\bf 41}, 395 (1978).

\bibitem{Rodriguez2018} I. Rodr'guez-Arias, I., M\"uller, M., Rosso, A. \& De Luca, A. An exactly solvable model for Dynamic Nuclear polarization. {\it Phys. Rev. B} {\bf 98}, 224202 (2018). 

\bibitem{DeLuca2015} De Luca, A. \& Rosso A. Dynamic nuclear polarization and the paradox of Quantum Thermalization. {\it Phys. Rev. Lett.} {\bf 115}, 080401 (2015). 

\bibitem{DeLuca2016} De Luca, A. Rodriguez-Arias, I., M\"uller, M.  \& Rosso, A. Thermalization and many-body localization in systems under dynamic nuclear polarization. {\it Phys. Rev. B} {\bf 94}, 014203 (2016). 

\bibitem{Mazurenko2017} 
Mazurenko, A., Chiu, C. S., Ji, G., Parsons, M. F., Kanasz-Nagy, M., Schmidt, R., Grusdt, F., Demler, E., Greif, D. \& Greiner, M. A cold-atom FermiÐHubbard antiferromagnet. {\it Nature}  {\bf 545}, 462 (2017).


\end{thebibliography}
\end{document}